\documentclass{article}
\usepackage{arxiv}

\usepackage{amsmath,amssymb,amsfonts}
\usepackage{algorithmic}
\usepackage{graphicx}
\usepackage{siunitx}
\usepackage{textcomp}
\usepackage{tabularx}
\usepackage{xcolor}
\def\BibTeX{{\rm B\kern-.05em{\sc i\kern-.025em b}\kern-.08em
    T\kern-.1667em\lower.7ex\hbox{E}\kern-.125emX}}
\usepackage[nohyperlinks, printonlyused, nolist]{acronym}

\acrodef{iot}[IoT]{Internet of Things}
\acrodef{iiot}[IIoT]{Industrial Internet of Things}
\acrodef{urllc}[uRLLC]{Ultra Reliable and Low Latency Communication}
\acrodef{xurllc}[xURLLC]{extreme Ultra Reliable and Low Latency Communication}
\acrodef{cpps}[CPPS]{cyber-physical production systems}
\acrodef{qos}[QoS]{Quality of Service}
\acrodef{nm}[NetEm]{network emulation}
\acrodef{nin}[NiN]{networks-in-network}
\acrodef{embb}[eMBB]{Enhanced Mobile Broadband}
\acrodef{kpi}[KPI]{key performance indicators}
\acrodef{dsm}[DSM]{Dynamic spectrum management}
\acrodef{cpf}[CPF]{control plane fabric}
\acrodef{dht}[DHT]{distributed hash table}
\acrodef{npn}[NPN]{non-public network}
\acrodef{ai}[AI]{artificial intelligence}
\acrodef{sm}[SM]{spectrum mnager}
\acrodef{sn}[SN]{sub-network}
\acrodef{snc}[SNC]{sub-network controller}
\acrodef{dt}[DT]{digital twin}
\acrodef{agv}[AGV]{automated guided vehicle}
\acrodef{pn}[PN]{private network}
\acrodef{nin}[NiN]{networks-in-network}
\acrodef{mrat}[multi-RAT]{multiple radio access technology}
\acrodef{ho}[HO]{handover}
\acrodef{gnb}[gNB]{Next-Generation NodeB}
\acrodef{ue}[UE]{user equipment}
\acrodef{nnpn}[NNPN]{nomadic non-public network}
\acrodef{sb}[SB]{spectrum broker}
\acrodef{csm}[CSM]{cognitive spectrum manager}
\acrodef{rrc}[RRC]{radio resource control}
\acrodef{plmn}[PLMN]{public land mobile network}
\acrodef{evm}[EVM]{error vector magnitude}
\acrodef{re}[RE]{resource element}
\acrodef{ndt}[NDT]{network digital twin}
\acrodef{cdu}[CDU]{central detection unit}
\acrodef{su}[SU]{sensing unit}
\acrodef{fpr}[FPR]{false positive rate}
\acrodef{gmm}[GMM]{Gaussian mixture model}
\acrodef{sdr}[SDR]{software-defined radio}
\acrodef{rss}[RSS]{received signal strength}
\acrodef{mqtt}[MQTT]{Message Queuing Telemetry Transport}
\begin{document}

%\bstctlcite{IEEEexample:BSTcontrol} % to control et al. etc

\title{Self-Healing 6G Networks-in-Network for Resilient Wireless Communication}

\author{
{Daniel~Lindenschmitt}\\
	Institute for Wireless Communication and Navigation\\
	RPTU Kaiserslautern-Landau\\
	\texttt{daniel.lindenschmitt@rptu.de} \\
	%% examples of more authors
    \And
{Anton~Schösser}\\
	Vodafone Chair Mobile Communications Systems\\
	Technische Universität Dresden\\
	\texttt{anton.schoesser@tu-dresden.de} \\
	%% examples of more authors
    \And
{Afnan~Aghai}\\
	Institute for Wireless Communication and Navigation\\
	RPTU Kaiserslautern-Landau\\
	\texttt{aghai@rptu.de} \\
	%% examples of more authors
    \And
{Philipp~Schulz}\\
	Vodafone Chair Mobile Communications Systems\\
	Technische Universität Dresden\\
	\texttt{philipp.schulz2@tu-dresden.de} \\\
	%% examples of more authors
    \And
{Gerhard~Fettweis}\\
	Vodafone Chair Mobile Communications Systems\\
	Technische Universität Dresden\\
	\texttt{gerhard.fettweis@tu-dresden.de} \\\
	%% examples of more authors
	\And
{Hans D.~Schotten}\\
	Institute for Wireless Communication and Navigation\\
	RPTU Kaiserslautern-Landau\\
	\texttt{schotten@rptu.de} \
}%

\maketitle

\begin{abstract}
Future 6G networks must manage increasingly dynamic radio environments in which multiple autonomous \acp{sn} share frequency resources and adapt to changing operating conditions. In such scenarios, interference from faulty devices or intentional jamming can  disrupt ongoing communications, making rapid and autonomous network adaptation essential. This demonstration presents a self-healing \ac{nin} architecture that closely integrates the detection of spectrum anomalies with dynamic spectrum management. A spectrum scanner continuously monitors the frequency spectrum and forwards detected anomalies to the DSM, which automatically identifies suitable frequency resources and reconfigures the affected SN. During the live demonstration, participants can initiate controlled disruptions and observe the entire adaptation process in real time, from anomaly detection to autonomous frequency reallocation and network recovery. The demonstrator illustrates how integrating spectrum monitoring and resource management into a single control loop can improve the resilience of future \ac{nin} implementations and demonstrates a practical approach to autonomous, spectrum-aware networking.
\end{abstract}

\keywords{
6G, Networks-in-Network, Anomaly Detection, Self-Healing, Spectrum Management, Resilience
}

\acresetall

\section{Introduction}
Wireless networks are becoming increasingly dynamic as more systems, applications, and services compete for limited spectrum resources. In future 6G deployments, this challenge could further be amplified by \ac{nin}, where several independent \acp{sn} share the same radio environment while operating independently. Under these conditions, maintaining reliable communication requires more than efficient spectrum allocation alone. The network must also be able to respond quickly to unexpected interference. In today's systems, spectrum monitoring and network management are often treated as separate tasks. An interference event may be detected within milliseconds, but network reconfiguration typically involves manual actions. This separation slows down the overall response and limits the network's ability to recover on its own.

In contrast to existing approaches that typically consider spectrum monitoring and spectrum management as separate functions, the proposed demonstrator directly couples anomaly detection with autonomous spectrum reconfiguration, allowing detection results to trigger reconfiguration of the affected \ac{sn}. Instead of only reporting that an interference has occurred, the system immediately starts the recovery process and migrates communication to an alternative channel whenever suitable spectrum is available before the radio link is fully blocked. Participants can observe the complete control loop during the live demonstration. An interfering noise signal is introduced into the radio environment, considered as an anomaly, which is detected by the monitoring system. The resulting spectrum decision is visualized and the affected \ac{sn} is automatically reconfigured. The demonstration therefore provides a practical example of how tightly coupled monitoring and spectrum management can support self-healing operation in wireless communication.

\section{Background}
\label{background}

Future 6G networks are expected to operate in highly dynamic environments where multiple independent networks coexist and share communication resources~\cite{Organic6G}. The \ac{nin} concept addresses this challenge by allowing several autonomous \acp{sn} to operate within an overlayer wireless network while preserving their individual control functions~\cite{10054381,sublayer}. Such architectures are particularly relevant for 5G or future 6G \acp{npn}, where communication requirements may differ considerably across applications and evolve over time. Therefore, efficient coordination of available spectrum becomes an essential requirement for a reliable network operation. \ac{dsm} has emerged as a key mechanism for coordinating spectrum access in these \ac{nin} environments. Depending on the deployment scenario, \ac{dsm} can assign spectrum resources to individual \acp{sn}, resolve conflicts between competing \acp{sn}, and adapt to changing radio conditions, such as those induced by spectrum anomalies. Previous work has explored centralized, distributed, and AI-assisted \ac{dsm} approaches for future 6G systems, demonstrating their potential to improve spectrum utilization while limiting mutual interference~\cite{11046311}. However, most existing solutions assume relatively stable operating conditions and treat spectrum management independently from anomaly detection or potential jamming. At the same time, wireless anomaly detection is gaining increasing attention as a method for identifying interference, malfunctioning devices, or intentional jamming. Recent approaches range from classical signal-processing techniques to advanced machine-learning-based approaches, capable of monitoring complex radio environments~\cite{rajendran2019unsupervised, sabanovic2024ai}. However, these methods are not fully exploited, often serving only as passive observers of the wireless spectrum. Integrating anomaly detection into network management could transform spectrum monitoring into a more active and integral part of wireless networks. This integration would not only enable automated responses to issues such as interference but also improve detection accuracy by incorporating meaningful context from the network management system. Such advancements are has the potential to significantly enhance the adaptability and resilience of wireless communication systems, unlocking new opportunities for \ac{dsm}.

\section{Anomaly Detection for Self-Healing Networks-in-Network}
\label{usecase}

Reliable operation in future \ac{nin} depends not only on detecting anomalies in the radio environment but also on responding to them before communication is significantly affected. For this reason, the demonstrated system combines spectrum monitoring with \ac{dsm} in a control loop. Rather than treating anomaly detection as a standalone monitoring task, the detection results are directly used to trigger network reconfiguration of the attached \acp{sn} whenever necessary.

Spectrum measurements are collected by a \ac{sdr} and processed in short observation intervals. For each interval, statistical features describing the received signal power are extracted and compared with values that have been collected earlier in a training phase under normal (i.e., in the absence of interfering signals) conditions. To separate normal channel behavior from anomalous observations, the training data (sequences of received power values) from normal operations are clustered using the $k$-means algorithm. If a sequence of observations significantly falls out of the known clusters, it is classified as an anomaly, and an alarm is raised to the \ac{dsm} entity. Because no anomaly data are used for training, the detector can identify a wide range of anomalies rather than being limited to specific patterns. K-Means was selected because it provides a lightweight unsupervised clustering method that can be executed efficiently on the edge hardware used in the demonstrator without requiring labeled training data. Accordingly, the detector is designed to distinguish normal spectrum behavior from localized interference events rather than to classify the origin of the interference.

The \ac{dsm} evaluates each reported anomaly together with the current spectrum allocation of the affected \ac{sn}. For the presented demonstrator, the decision process primarily considers the measured \ac{rss}, since this metric is continuously available from the spectrum scanner and can be evaluated without introducing additional signaling overhead. The modular architecture allows additional \ac{qos} indicators, such as latency or throughput, to be incorporated into future spectrum selection strategies. If the \ac{rss} is expected to degrade, the \ac{dsm} selects a suitable alternative channel and initiates an autonomous spectrum reconfiguration. In this way, anomaly detection becomes an integral part of the resource management process rather than an isolated monitoring function. The developed architecture further separates sensing, anomaly detection, and spectrum management, making it possible to replace the current detection algorithm or integrate distributed sensing nodes in future implementations~\cite{10741442, lindenschmitt2025adaptive}.

\section{Demo Setup}
\label{demo}

\begin{figure}
    \centering
    \includegraphics[width=0.9\linewidth]{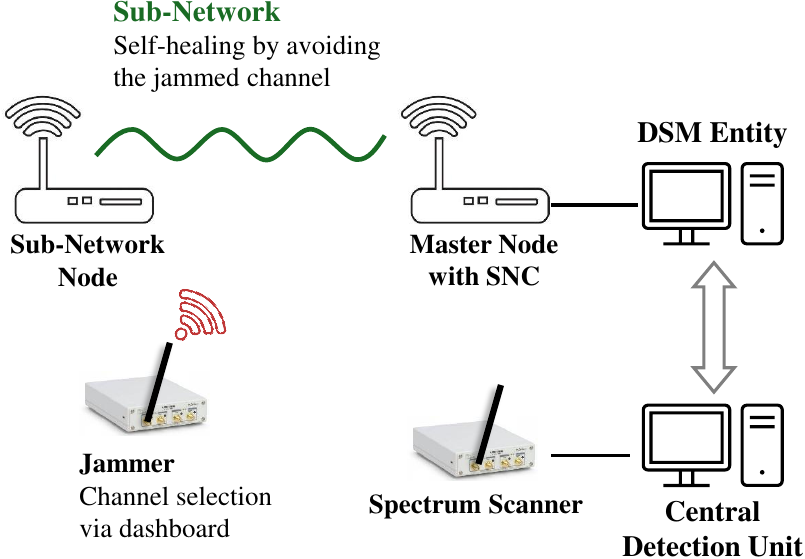}
    \caption{System architecture of the demonstration}
    \label{fig:setup-schematic}
\end{figure}

The setup consists of the components illustrated in Figure~\ref{fig:setup-schematic}: 

\begin{itemize}
    \item \emph{Spectrum scanner} for radio environment monitoring,
    \item \emph{\ac{cdu}} for analyzing monitored values and anomaly detection,
    \item \emph{\ac{dsm} entity} for processing feedback sent by \ac{cdu},
    \item \emph{\ac{sn}} to adapt to changed radio conditions based on recommendations by the \ac{dsm}, and 
    \item \emph{Jammer} to cause artificial anomalies to the setup.
\end{itemize}

The radio environment is continuously observed by a spectrum scanner which is realized by an \ac{sdr} (\emph{USRP B210}). It provides sweeping spectrum information with \SI{20}{MHz} bandwidth within the frequency range from \num{3.7} to \SI{3.8}{GHz} to the anomaly detection algorithm operating in the \ac{cdu} on a \emph{LattePanda Sigma}. Instead of searching for predefined interference patterns, the detector identifies deviations from the expected spectrum behavior and reports them to the \ac{dsm}. This keeps the detection process lightweight while remaining flexible enough to react to different types of interference without requiring prior knowledge of the signal source. Reporting and data exchange is realized via a \ac{mqtt} broker, where all attached components can publish their values and subscribe to the respective ones. The advantage of implementing via \ac{mqtt} is a fast and continuous data exchange, which still supports the modular approach of this demonstration. Only compact status information and control commands are exchanged between the functional components. As a result, the communication overhead introduced by \ac{mqtt} remains low compared to the application traffic carried by the \ac{sn}.

The \ac{dsm} is also executed on a \emph{LattePanda Sigma} and subscribed to regular updates from the \ac{cdu}. Based on the reported anomaly, the \ac{dsm} determines whether the current spectrum allocation is still suitable for the affected \ac{sn}. The \ac{sn} itself consists of two network nodes, where one is acting as a master and the other as a slave node. The master node is triggering the network parameter adaptation based on recommendations by \ac{dsm}. If continued operation on the active channel is no longer feasible, an alternative spectrum resource is selected, and the \ac{sn} is reconfigured accordingly. If available,  remaining \acp{sn} continue to operate without modification, limiting the impact of the interference to the affected communication domain.

The system also features a jammer that is able to generate interference in the system’s frequency range as needed. This interference is then detected ad hoc by the \ac{cdu}, which initiates further actions via the \ac{dsm} as necessary. Based on the \ac{dsm}'s recommendations, the \ac{sn} can switch to one of several available center frequencies,  before communication quality degrades significantly. The demonstrator considers localized narrowband interference affecting the operating channel of a single \ac{sn}. Such interference may originate from unintentional spectrum usage or intentional jamming. Wideband interference covering the complete available spectrum, simultaneous interference affecting multiple \acp{sn}, or failures of the \ac{dsm} itself are outside the scope of the current demonstration. To illustrate more clearly how the self-healing process works, only two center frequencies at \SI{3.71}{GHz} (channel~1) and \SI{3.75}{GHz} (channel~5) were used for this demo.

The overall architecture separates sensing, decision-making, and spectrum management into independent functional blocks connected via clearly defined interfaces. This modular structure simplifies future extensions, such as integrating distributed sensing nodes or replacing the current detection method with more advanced approaches, while leaving the overall control process unchanged.

\section{Demo Walkthrough}
\label{walk}

\begin{figure}
    \centering
    \includegraphics[width=\linewidth]{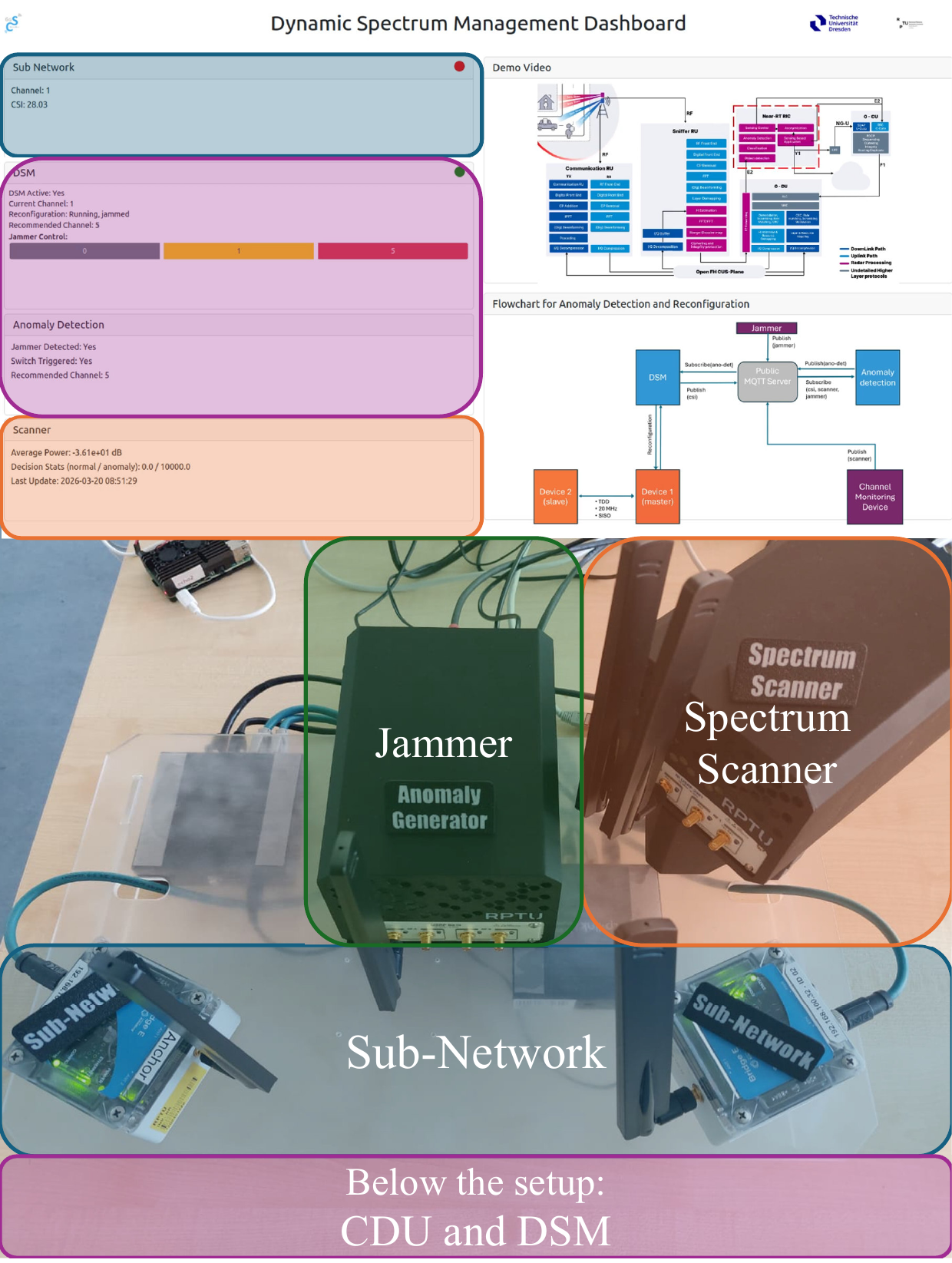}
    \caption{Overview of demonstration setup}
    \label{fig:setup-hardware}
\end{figure}

During the demonstration, interference is intentionally introduced on the operating channel of one \ac{sn}. The resulting anomaly is detected automatically, after which the \ac{dsm} initiates the spectrum reconfiguration and migrates the affected communication to an alternative available channel. The demonstrated recovery represents the complete control loop from anomaly detection to successful network reconfiguration, including the exchange of control information between the anomaly detector and the \ac{dsm}. The complete workflow can be observed in real time and illustrates how anomaly detection and \ac{dsm} complement each other to provide autonomous self-healing capabilities for \ac{nin} deployments. The presentation is divided into:
\begin{itemize}
    \item \emph{Jammer Interaction:} Participants can interact with the jammer via the system's dashboard. They can choose to have no active jammer, or whether the jammer should interfere with channel 1 or channel 5. To ensure that no wireless networks are actually disrupted, the jammer is designed as a virtual jammer for the demonstration. 
    \item \emph{System Behavior Visualization:} Participants can monitor the jammer's impact as well as the current status of the frequency spectrum and the \ac{sn} being monitored directly through the dashboard. If the jammer is activated, the devices are reconfigured via the \ac{cdu} and the \ac{dsm}, a process that is also visualized on the dashboard.
\end{itemize}

%Additional material, as a screen-recording of the dashboard during the demonstration, can be found here: \emph{seafile.rlp.net/f/c4011ed9fa4e4211b1ac/}

\section{Conclusion and Future Work}
\label{concl}
The presented demonstrator illustrates how spectrum anomaly detection and \ac{dsm} can be combined to realize autonomous self-healing behavior in future \ac{nin} deployments. The focus of the demonstration is on validating the functionality of the complete control loop under controlled laboratory conditions rather than on evaluating large-scale deployments or optimizing individual processing stages. Future work will extend the current demonstrator by incorporating distributed sensing, \ac{qos}-aware decision metrics, more sophisticated anomaly detection methods, and evaluations in larger \ac{nin} deployments with multiple interacting \acp{sn} and concurrent interference events.

%The presented demonstrator shows how anomaly detection and \ac{dsm} can work together to support self-healing behavior in a future 6G \ac{nin} environment. By connecting spectrum monitoring directly to the network adaptation process, the system is able to react to interference without manual intervention, enabling a fast and autonomous solution for resilient spectrum management. The current implementation represents a first step toward autonomous self-healing networks. Future work will focus on extending the concept to distributed sensing, implementation of more sophisticated detection methods, and larger, more dynamic \ac{nin} deployment in order to improve scalability, support mobility, and enable more robust detection strategies.

\section*{Acknowledgment}
The authors acknowledge the financial support by the German \textit{Federal Ministry of Research, Technology and Space (BMFTR)} within the projects Open6GHub+ \{16KIS2406\} and 6G-CampuSens \{16KISK207 and 16KISK208\}. 

\bibliographystyle{IEEEtran}
%{
\bibliography{references}
%}
\end{document}